\documentclass[12pt]{iopart}
\usepackage{iopams}
\usepackage{graphicx}
\usepackage[caption=false,farskip=0pt,labelfont={bf}]{subfig}
\usepackage{dcolumn}
\usepackage{bm}
\usepackage{paralist}
\usepackage{cite}

\begin{document}

\title{Exact results of dynamical structure factor of Lieb-Liniger model}

\author{Run-Tian Li$^{1}$}

\author{Song Cheng$^{2}$}
\ead{scheng@csrc.ac.cn}

\author{Yang-Yang Chen$^{1}$}
\ead{chenyy@nwu.edu.cn}

\author{Xi-Wen Guan$^{3,4,5}$}
\ead{xwe105@wipm.ac.cn}

\address{$^1$Institute of Modern Physics, Northwest University, Xi'an 710069, China}
\address{$^2$Beijing Computational Science Research Center, Beijing 100193, China }
\address{$^3$State Key Laboratory of Magnetic Resonance and Atomic and Molecular Physics, Wuhan Institute of Physics and Mathematics, Chinese Academy of Sciences, Wuhan 430071, China}
\address{$^4$Department of Theoretical Physics, Research School of Physics and Engineering, Australian National University, Canberra ACT 0200, Australia}
\address{$^5$Peng Huanwu Center for Fundamental Theory, Xi'an 710069, China}

\date{\today}


\begin{abstract}
The dynamical structure factor (DSF) represents a measure of dynamical density-density correlations  in a quantum  many-body system.
Due to the complexity of many-body correlations and quantum  fluctuations in a system of an infinitely large  Hilbert space, such kind of dynamical correlations often impose a big theoretical challenge.
For one dimensional  (1D) quantum many-body systems,  qualitative predictions of dynamical response functions are usually  carried out by using the Tomonaga-Luttinger liquid (TLL) theory.
In this scenario,  a  precise evaluation of the DSF for a 1D quantum system with arbitrary interaction strength  remains a formidable task.
In this paper, we use the form factor approach based on algebraic  Bethe ansatz theory to calculate precisely the DSF of Lieb-Liniger model with an arbitrary interaction strength at a large scale of particle number.
We find that the DSF for a system as large as 2000 particles enables us to depict precisely its line-shape from which the power-law singularity  with corresponding exponents in the vicinities of spectral thresholds naturally emerge.
It should be noted that, the advantage of our algorithm promises an access to the threshold behavior of dynamical correlation functions, further confirming the validity of nonlinear TLL theory besides Kitanine {\em et. al.} 2012 {\em J. Stat. Mech.} P09001.
Finally we discuss a comparison of results with the results from  the ABACUS method  by J.-S. Caux 2009 {\em J. Math. Phys.} {\bf 50} 095214 as well as from the strongly coupling expansion by Brand and Cherny 2005 {\em Phys. Rev.} A {\bf 72} 033619.
\end{abstract}




\noindent

\section{Introduction}

A measurement on dynamical correlations of many-body systems  can be implemented by inserting into a probe, and then its scattering with constituents produces density fluctuations and tunneling rates, building on which the  information one may obtain is   the dynamical structure factor (DSF)  and spectral function (SF) \cite{Pines}.
From a theoretical perspective, the DSF of a many-body system in three-dimension can be figured out by utilizing Green's function method \cite{Abrikosov}.
A stark contrast to this situation is the study of dynamic correlated properties in one-dimension, wherein the reduced dimensionality and enhanced quantum fluctuation induce a non-perturbative characteristic and thus put forward new challenges.
In this scenario,  a variety of fruitful methods have been employed such as conformal field theory (CFT) \cite{Francesco}, bosonization and its extension \cite{Cazalilla2004,Cazalilla,Giamarchi,Imambekov2012,Gogolin}, quantum integrable theory \cite{QISM,Franchinni} and density matrix renormalization group (DMRG) method \cite{Schollwock} etc.
However, CFT and bosonization focus on the universal properties at a low energy  and lack of a full-picture illustration for dynamical  correlation functions.
Whereas the DMRG is more suitable for the  lattice systems.

Even for the arguably simplest  spinless  Bose gas  \cite{Lieb}, the exact evaluation of correlation functions is hard to tackle once the interaction is present, despite at a certain  limit the wave function of the Tonks-Girardeau (TG) gas is known \cite{Girardeau}.
In the TG gas, the repulsive boson is a reminiscent of non-interacting fermions due to the dynamic interaction, allowing a quite interesting and subtle correspondence between them.
This Bose-Fermi map \cite{Girardeau} offers a reliable access to the correlated properties of TG gases either of continuum or on a lattice, such as the DSF, one-body density matrix, one-particle dynamical correlation function, and spectral function (SF) etc \cite{Lenard,Tracy,Buljan,Minguzzi,Colcelli2018a,Colcelli2018b,Colcelli2020}.
A little bit away from the TG limit, the generalization of this map works for strongly interacting Bose gases as well, the DSF of ground state and finite temperatures of the Lieb-Liniger model are derived based on a pseudopotential Hamiltonian \cite{Brand2005,Cherny}.
Turing to generic interaction strength, it is totally different and difficult.
Having obtained some power-law behaviour on basis of bosonization and its extension \cite{Cazalilla,Giamarchi,Imambekov2012,Cazalilla2004,Gogolin},
the exact evaluation of DSF is still in great demand.

The Lieb-Liniger model describes $N$ spinless bosons with a contact  interaction   in a line \cite{Lieb}.
As one of the simplest quantum integrable models, it has been extensively  studied  on a variety of aspects, including  thermodynamic properties and quantum criticality \cite{Jiang,Guan2011}, quantum interaction quench \cite{Nardis2014,Buljan2008,Muth2010,Andrei,Mossel2012,Kormos2013,Kormos2014}, static correlation functions \cite{Shlyapnikov,Nandani,Xu2015} and dynamical correlation functions \cite{Brand2005,Cherny,Caux2006,Caux2007,Caux2009,Caux2023,Panfil,Ours_1,Ours_2,Granet,Rosi2022,Rosi2023}  etc.
A quantum integrable system amenable to the Yang-Baxter equation does promise exactly solvability, however  it is hardly to carry out an investigation into correlations naively from the Bethe wavefunction at a many-body level \cite{QISM,Franchinni}.
On the other hand, the algebraic Bethe ansatz (ABA) provides a way to formulate a determinant representation for the form factor of a physical observable through rapidities of Bethe Ansatz equations (BAEs) \cite{QISM,Slavnov1997,Slavnov,Korepin}.
The first successful implementation of transforming these sophisticated formulas into a useful  form comes from the algebraic Bethe Ansatz-based Computation of Universal Structure factors (ABACUS) method developed by J.-S. Caux \cite{Caux2009}.
In this way, the DSF \cite{Caux2006} and one-particle dynamic correlation \cite{Caux2007} of ground state were  calculated for a system consisting of  100 and 150 particles, respectively.
However, due to the limitation of  scanning algorithm and data treatment, the ABACUS method is not suitable  to treat the threshold behavior of dynamical correlation functions \cite{Caux2009}.

In this scenario,  we develop  an efficient  algorithm to compute the dynamical correlation functions of the Lieb-Liniger model with an arbitrary interaction strength by sophisticatedly counting  the relevant intermediate states in the Hilbert space  \cite{Ours_1,Ours_2}.
Using this algorithm together with properly defining a set of four quantum numbers  in classifying the `relative excitations' over a reference state, we obtain the  DSF with so far the highest  accuracy and consequently its threshold behavior.
In particular, the line-shape and the singular behavior of DSF close to the spectral threshold are calculated at a very large scale $N=2000$.
The emergent exponents for the edge singularity are numerically extracted  and  essentially confirm the validity of nonlinear Tomonaga-Luttinger (TLL) theory.
We further compare our results with  the ones obtained from the ABACUS  \cite{Caux2009} and other theoretical results in strongly interacting region \cite{Brand2005}.
It turns out that our algorithm offers an excellent access to the line-shape and the threshold power-law of the DSF at a significantly many-body scale, see the following discussion on Figures~2 - 9.

\section{Model and DSF}
\begin{figure}[ht]
\centering
\includegraphics[width=0.8\textwidth]{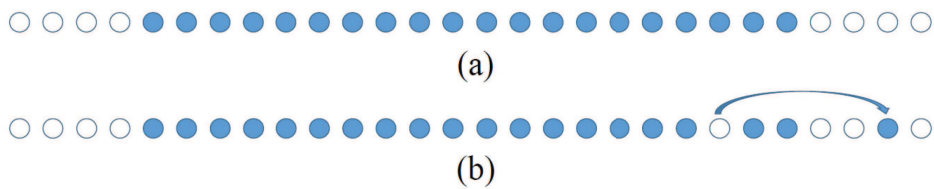}
\caption{(a) displays the configuration of QNs in ground state,  showing an analog of the Fermi sea.
For a $N$-particle system, the ground state QNs lie between $-(N-1)/2$ and $(N-1)/2$ consecutively.
(b) shows a typical  example of $1$-pair of particle-hole (p-h) excitation.}
\label{excitation}
\end{figure}

The Hamiltonian of the Lieb-Liniger model reads
\begin{equation}\label{hamiltonian}
H = - \sum_{i=1}^{N} \frac{\partial^2}{\partial x_i^2} + 2c \sum_{i>j}^{N} \delta \left( x_i - x_j \right),
\end{equation}
where $c>0$ ($c<0$) specifies the repulsive (attractive) interaction.
Here we only consider  the repulsive case.
For later convenience in calculation, we define the particle density $n=N/L$ and a dimensionless interaction parameter
$\gamma = c/n$.
Tuning this parameter $\gamma$  in the whole repulsive region, i.e. $0 \rightarrow \infty$, the ground state of system varies from quasi-condensate to TG gas.
Substituting the Bethe wavefunction into the Hamiltonian, one obtains a set of  transcendental equation, i.e. the Bethe ansatz equations (BAEs) \cite{Lieb}
\begin{equation}\label{BAE}
\lambda_j + \frac{1}{L} \sum_{k=1}^{N} \theta \left( \lambda_j - \lambda_k \right) = \frac{2\pi}{L} I_j, \quad j = 1, \dots, N,
\end{equation}
where $\theta(x)=2\arctan(x/c)$, $\lambda_j$ and $I_j$ are respectively the rapidity and  its corresponding quantum number (QN).
A set of $\{ I_j \}$ uniquely determines a quantum state (represented by a set of $\{\lambda_j\}$), vice versa.
Those QNs taking integer and half-integer depends on the parity of $N$, see Figure~\ref{excitation}.
The total momentum and energy of the system can be expressed in forms of rapidities
\begin{equation}\label{PE}
P_{\{\lambda\}}=\sum_{j=1}^{N} \lambda_j, \qquad E_{\{\lambda\}}=\sum_{j=1}^{N} \lambda_j^2.
\end{equation}

It is convenient to describe the eigenstates by employing the configurations of QNs.
As is sketched in Figure~\ref{excitation}, the ground state is depicted by a Fermi sea-like distribution, over which pairs of particle-hole (p-h) excitations simply generate excited states.
In principle, an arbitrary eigenstate of the Hilbert space can be accessed by manipulating a series of pairs of p-h excitations.

The dynamical structure factor \cite{LP} is defined by
\begin{equation}\label{DSF_def}
S (k,\omega) = \frac{L}{2\pi} \int_{0}^{L} \textmd{d}x  \int_{-\infty}^{\infty} \textmd{d}t \, e^{\rmi (kx-\omega t)} \langle \rho(x,t) \rho(0,0) \rangle,
\end{equation}
where $\rho(x,t)$ is the density operator and $\langle \cdots \rangle$ means expectations taken over the state of our interest.
In our realistic calculations, however, the finite temperature correlation is not carried out naively from this  definition.
From the macroscopic viewpoint, any equilibrium state of quantum integrable systems can be described by the thermodynamic Bethe ansatz (TBA) equations \cite{note1}.
It is not an eigenstate of the system,  but a mixture of microscopic states sharing the same macroscopic description of a total number $e^{\mathcal{S}}$,  where $\mathcal{S}$ is the entropy.
The expectation over arbitrary one of these eigenstates gives the same result.
Therefore the problem at finite temperature now is reformulated at a similar level to the ground state one.
Under  this circumstance, what we need is nothing but a discredited solution to the TBA.
Below we call this eigenstate of our interest as reference state which can be either the ground state or a microscopic state bearing the macroscopic description for the equilibrium state at nonzero temperatures, see Figures~\ref{tag_T_Finite} of Section Method for a simple example.

The spectral expression is obtained by inserting the completeness equality
$\sum _{|\{\mu\}\rangle} | \{\mu\} \rangle \langle \{ \mu \} | /\langle \{\mu\} | \{ \mu\} \rangle =1$ into Equation~\ref{DSF_def}
\begin{equation}\label{spectral_form}
S \left( k,\omega \right) = L^2  \sum_{\{\mu\} } \frac{ \|  \langle \{\mu\} | \rho(0,0)  |  \{\lambda\} \rangle \|^2 }{\|  \{\lambda\} \|^2  \| \{\mu\} \|^2}  \,
\delta_{k,P_{\mu,\lambda}}  \delta\left(\omega-E_{\mu,\lambda} \right),
\end{equation}
where $\delta_{m,n}$ and  $\delta(x)$ are the Kronecker- and Dirac-delta functions,  respectively.
For simplicity, we denote $O_{ \mu,\lambda} \equiv O_{\{\mu\}}- O_{\{\lambda\}}$ with $O=E$ or $P$.
$\langle \{ \mu \} | \rho(0,0)  | \{ \lambda\} \rangle $ is called \textit{form factor} and $\| \{ \nu \} \|^2$ is the norm square of  eigenstate, both of which can be transformed into determinants with entries represented by rapidities \cite{Slavnov}.
The summations for intermediate states $| \{ \mu \} \rangle$ is assumed to be over the whole Hilbert space, yet impossible in practice.
The reasonable solution is to incorporate states with significant contributions to DSF as many as possible, before satisfying the required accuracy.
The saturation is readily checked by the $f$-sum rule  \cite{LP}
\begin{equation}\label{sum_rule}
\int_{-\infty}^{\infty} \textmd{d}\omega\, \omega S(k,\omega) = N k^2.
\end{equation}
It should be noted that the DSF possess nonzero value at negative energy only if at finite temperatures, and for simplicity a normalized   form of above $f$-sum rule is used later on.

\section{Methods}
It is obvious that the key to evaluating dynamical correlation functions through form factor approach is how to efficiently and quickly find the essential  states in the process of navigating Hilbert space.
To this end, recently we have developed an algorithm suitable for calculating various dynamical correlation functions, such as single particle Green's functions and one particle density matrix at both zero and finite temperatures \cite{Ours_1,Ours_2}.
The idea involves  two different aspects: on one side, as is explained before, the correlation function at a thermodynamic equilibrium state is transformed into a similar problem to that for ground state, i.e. expectation over an eigenstate satisfying  the TBA  equations;
on the other side, the most relevant states for calculating the  form factors of a local observable ought not to be much different to the reference state in  the perspective of QNs configuration.

\begin{figure}[htbp]
  \centering
  \includegraphics[width=1.1\textwidth]{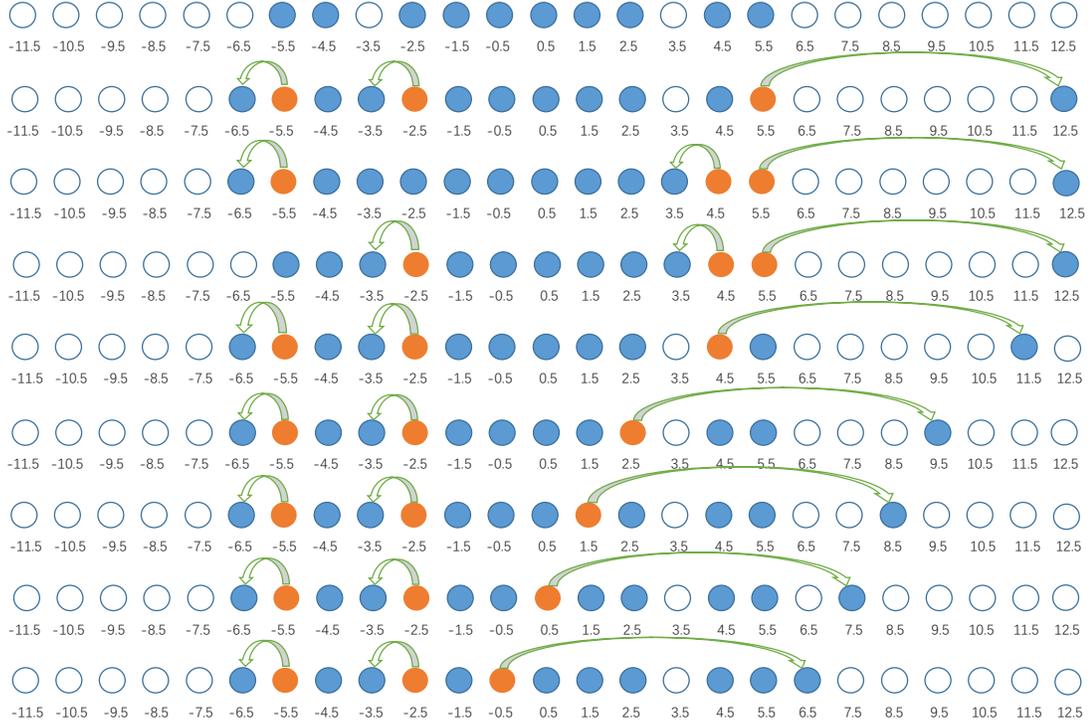}
  \caption{The first row  is the reference state of the  system with $N=L=10$, interaction $c=4$ at temperature $T=3$.
  The balls mean vacancies of  QNs,  the blue (white) ones are occupied (un-occupied) vacancies and they  stand for  particles (holes).
  The arrow represents a `relative excitation', and the orange balls indicate the particle taking part in that `excitation'.
  The following eight `excited states' over this reference state belong to the set of tags $(P_\mathrm{m},N_\mathrm{p},P_\mathrm{l},N_\mathrm{l})=(5, 3, 2, 2)$.
  Their total contributions to the DSF is $9.4639 \times 10^{-5}$ as being summarized in Table~\ref{tab}, and also  Figure~\ref{treeT}.
  }
  \label{tag_T_Finite}
\end{figure}

\begin{figure}[htbp]
  \centering
  \includegraphics[width=1.1\textwidth]{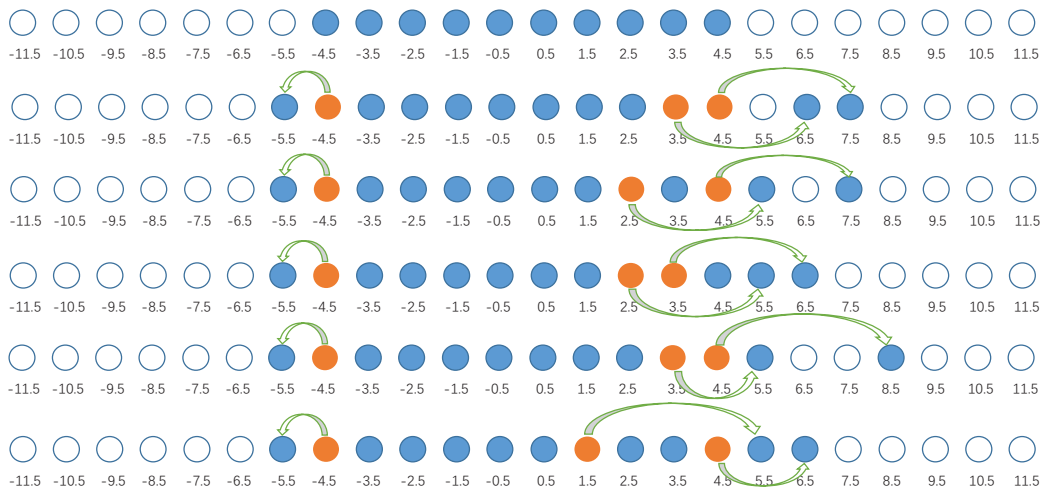}
  \caption{The first row is the ground state of a system with $N=L=10$ and interaction $c=4$.
  The following are the excited states belonging to the set of tags $(P_\mathrm{m}, N_\mathrm{p}, P_\mathrm{l}, N_\mathrm{l})=(5, 3, 1, 1)$.
  Their contribution  to the weights of the sum rule for the DSF is $7.7379 \times 10^{-5}$,
  also see the case at the ground state showing in  Table~\ref{tab_T0} and Figure~\ref{tree}.
  The notations for the different color balls are the same as that in Figure~\ref{tag_T0}.
   Obviously,  the `relative excitations' here reduce to the conventional p-h excitations as being shown by Figure~\ref{excitation}.}
  \label{tag_T0}
\end{figure}

In light of the prospectively unified description for the reference state, one has to abandon the conventional understanding of  p-h excitations shown in Figure~\ref{excitation}.
We hence define arbitrary re-distributions of QNs away from the reference state as the `relative excitations' over it.
Before moving on to the details of our algorithm, we give an example of the reference state and `relative excitations'  at finite temperatures in   Figure~\ref{tag_T_Finite}.
 There we illustrate the reference state for a thermodynamic equilibrium state of interaction $c=4$ and temperature $T=3$  with  a small size of $N=L=10$.
The first row corresponds to the  reference state, i.e. one of the eigenstates satisfying the macroscopic descriptions of the TBA equations  \cite{note1}.
 The following rows in  the figure show the states produced by three `relative excitations', and their classification will be discussed later on.
 Each arrow clearly marks the direction and step-length of the `relative' p-h excitation.
 The blue (white) balls stand for the positions of QNs of the particles (holes) and the orange ones  indicate the particles making up `relative excitations'.
With this setting, the conventional p-h excitation is clearly  a special case of the `relative excitation'.
This is demonstrated by the Figure~\ref{tag_T0}, where the first row is exactly the ground state of the same system with $N=L=10$ and interaction $c=4$, while the rest rows show states produced by three excitations as well.

Let us first  consider how to classify the excited states over a reference state.
For  this purpose, we introduce a set of four tags $(P_\mathrm{m},N_\mathrm{p},P_\mathrm{l},N_\mathrm{l})$.
They are four non-negative integers, $P_\mathrm{m}=k*L/2\pi$ is employed to specify the value of momentum of an excited state, $N_\mathrm{p}$ is the number of particles involved in relative excitations, $N_\mathrm{l}<N_\mathrm{p}$ is the number of particles jumping leftward, and $P_\mathrm{l} \geq N_\mathrm{l}$ is the sum of leftward step-length for all $N_\mathrm{l}$ particles in units of $2\pi/L$.
They can be seen as a set of quantum numbers to describe the `relative excitations' over a reference state.
In this way, the Hilbert space is separated into a large number of subspaces, and one may find the most relevant states by a proper choice of the sets of tags.
It should be stressed  that the strategy of our navigation of states is rather different to the ABACUS \cite{Caux2009} in several aspects.
The value of momentum is chosen as the first tag, making an efficient way of counting states for computing the line-shape of dynamic correlation functions.
We define the `relevant excitations' and use their number $N_\mathrm{p}$ instead of the conventional p-h excitation number.
On account of the mirror-symmetric distribution of DSF with respect to energy axis, we make use of four tags of  $(P_\mathrm{m},N_\mathrm{p},P_\mathrm{l},N_\mathrm{l})$  instead of energy as  a  criteria for cut-off.
All these are essentially different from the counting manner in  \cite{Caux2009}.
Since this partition based on sets of tags depends on the reference state, it is efficient to  accelerate the search of those relevant states that make non-negligible sum-rule weights.
This method is  especially efficient  for finite temperature situation.

\begin{figure}[htbp]
\centering
\includegraphics[width=1.1\textwidth]{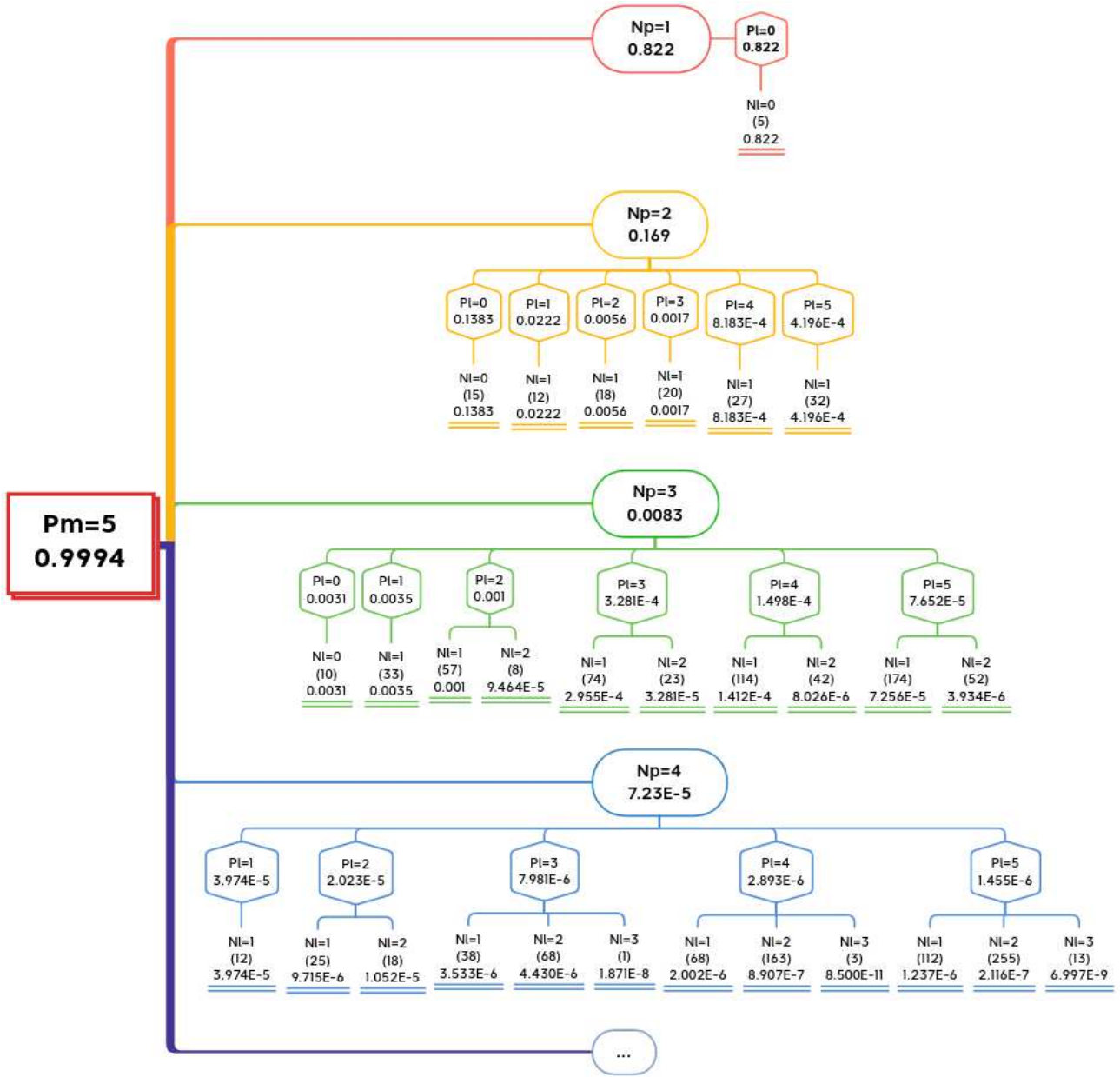}
\caption{The sum rule  weights associated with the  data given in Table~\ref{tag_T_Finite} for   the DSF  $S(k_\mathrm{F},\omega)$ at a finite temperature with $N=L=10$, interaction $c=4$ and temperature $T=3$.
Each set of tags $(P_\mathrm{m},N_\mathrm{p},P_\mathrm{l},N_\mathrm{l})$ attributes to  the roots of the sum rule  tree.
The number in the parentheses at the end root  for each  $N_\mathrm{l}$ is the number of states belonging to the set of tags.
The weight of the  $f$-sum rule of each branch is given level by level, i.e. containing a sum of all  branches of next level.
The eight states of tags $(5, 3, 2, 2)$ are also  shown in Figure~\ref{tag_T_Finite}.
}.
\label{treeT}
\end{figure}

\begin{figure}[htb]
\centering
\includegraphics[width=0.7\textwidth]{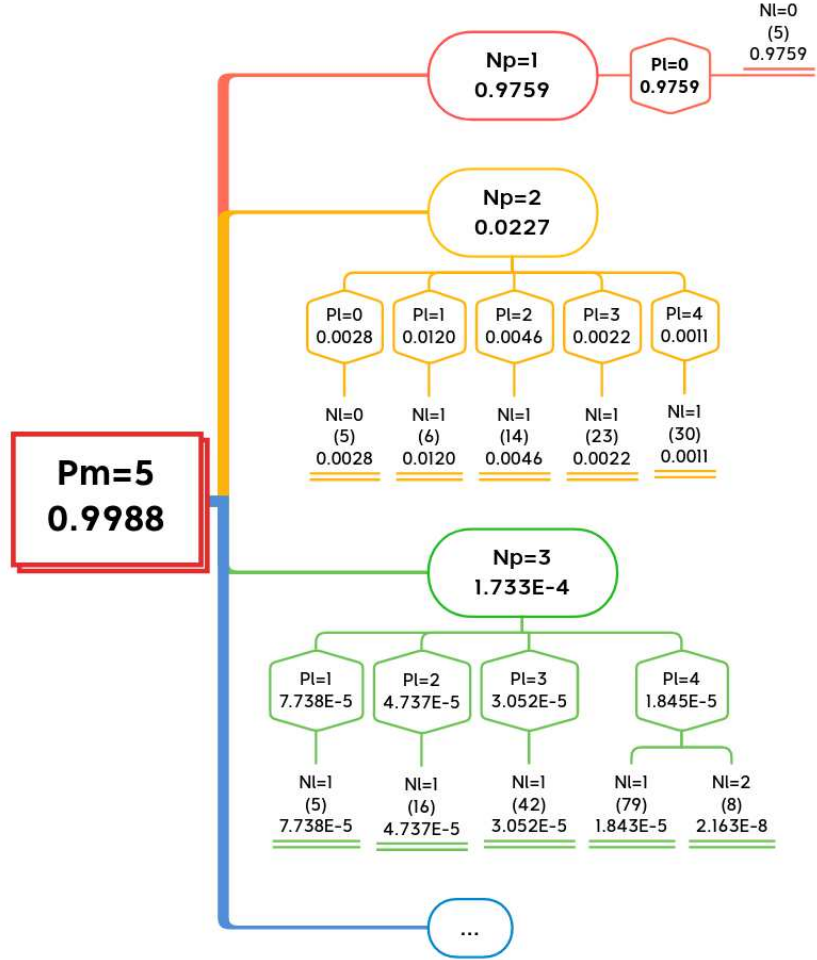}
\caption{The sum rule  weights associated with the  data given in Table~\ref{tag_T0} for  DSF  $S(k_\mathrm{F},\omega)$ in  the ground state of the system with $N=L=10$ and interaction $c=4$.
Each set of tags $(P_\mathrm{m},N_\mathrm{p},P_\mathrm{l},N_\mathrm{l})$ attributes to  the roots of the sum rule  tree.
The number in the parentheses at the end root  for each  $N_\mathrm{l}$ is the number of states belonging to the set of tags for the ground state.
The weight of the  $f$-sum rule of each branch is given level by level, i.e. containing a sum of all  branches of next level.
The eight states of tags $(5, 3, 1, 1)$ are also  shown in Figure~\ref{tag_T0}.}
\label{tree}
\end{figure}

We give  an example, the eight excited states belonging to the set of tags $(P_\mathrm{m},N_\mathrm{p},P_\mathrm{l},N_\mathrm{l})=(5, 3, 2, 2)$ with respect to reference state determined by interaction $c=4$ and temperature $T=3$, see Figure~\ref{tag_T_Finite}.
A similar example for ground state is displayed by Figure~\ref{tag_T0} with $(P_\mathrm{m}, N_\mathrm{p}, P_\mathrm{l}, N_\mathrm{l})=(5, 3, 1, 1)$.
For the sake of clarity, below we sketch the construction of the five excited states in Figure~\ref{tag_T0}.
There are $N_\mathrm{p}=3$ orange balls, one of them jump leftward with a step-length $1$, and the two particles jumping rightward need to move a total sum of step-length $P_\mathrm{m}+P_\mathrm{l}=6$ for compensation of the leftward excitations.
Then the division of $6$ into a sum of $2$ positive integers alas $3+3$, $2+4$ and $1+5$ leaves only five available arrangements.
For the  number of states belonging to different sets of tags, one may refer to Figure~\ref{treeT} and Table~\ref{tab} (Figure~\ref{tree} and Table~\ref{tab_T0}), illustrating  an example of finite temperature excitations (excitation over ground state similarly).

\begin{table*}
\caption{\label{tab} The sum rule weights  of the  DSF at $k= k_F$ of the system with $N=L=10$, interaction strength $\gamma=4$ and temperature $T=3$. The X-sum rule is the total spectral weights of the states under tag X,  and the  $N_\mathrm{p}$-sum rule specifies  the contributions of different pairs relative p-h excitations.
The tree structure of this sample is shown in Figure~\ref{treeT} and the states belonging to the set of tags $(5, 3, 2, 2)$ are demonstrated in Figure~\ref{tag_T_Finite}.
}
\lineup
  \begin{tabular}{c | c | c | c | c c c c c}
  \br
  \multicolumn{4}{c}{\textbf{Tags}} & \textbf{Number of} & \multicolumn{4}{c}{\textbf{Sum Rule}}\\
   \multicolumn{4}{c}{} & \textbf{ States} & \multicolumn{4}{c}{}\\
  \mr
  $P_\mathrm{m}$ & $N_\mathrm{p}$ & $P_\mathrm{l}$ & $N_\mathrm{l}$ &  & $N_\mathrm{l}$-sum rule &  $P_\mathrm{l}$-sum rule & $N_\mathrm{p}$-sum rule & $P_\mathrm{m}$-sum rule \\
  5  & 1  &  0  &  0  & 5         & $0.8220$         & $0.8220$                  & $0.8220$           &  $0.9994$    \\
  5  & 2  &  0  &  0  & 15       & $0.1383$         & $0.1383$                  & $0.1690$           &  {} \\
  5  & 2  &  1  &  1  & 12       & $0.0222$         & $0.0222$                  & {}                       &  {} \\
  5  & 2  &  2  &  1  & 18       & $0.0056$         & $0.0056$                  & {}                       &  {} \\
  5  & 2  &  3  &  1  & 20       & $0.0017$         & $0.0017$                  & {}                       &  {} \\
  5  & 2  &  4  &  1  & 27       & $8.1834 \times 10^{-4}$         & $8.1834 \times 10^{-4}$                  & {}                       &  {} \\
  5  & 2  &  5  &  1  & 32       & $4.1960 \times 10^{-4}$         &$4.1960 \times 10^{-4}$                   & {}                       &  {} \\
  5  & 2 & $\vdots$   & $\vdots$   & $\vdots$             & $\vdots$      &   $\vdots$                            &  {}                                               &  {} \\
  5  & 3  &  0  &  0  & 10       & 0.0031                                     & 0.0031                                              &  0.0083                                     & {} \\
  5  & 3  &  1  &  1  & 33       & 0.0035                                     & 0.0035                                              &  {}                                             &  {} \\
  5  & 3  &  2  &  1  & 57       & 0.0010                                     & 0.0010                                              &  {}                                             &  {} \\
  5  & 3  &  2  &  2  & 8         & $9.4639 \times 10^{-5}$        & {}                                                      &  {}                                             &  {} \\
  5  & 3  &  3  &  1  & 74       & $2.9547 \times 10^{-4}$        & $3.2809 \times 10^{-4}$                  &  {}                                             &  {} \\
  5  & 3  &  3  &  2  & 23       & $3.2809 \times 10^{-5}$        & {}                                                      &  {}                                            &  {} \\
  5  & 3  &  4  &  1  & 114     & $1.4178 \times 10^{-4}$        & $1.4980 \times 10^{-4}$                  &  {}                                            &  {} \\
  5  & 3  &  4  &  2  & 42       & $8.0263 \times 10^{-6}$        & {}                                                      &  {}                                               &  {} \\
  5  & 3  &  5  &  1  & 174    & $7.2585 \times 10^{-5}$         & $7.6523 \times 10^{-5}$                 &  {}                                                &  {} \\
  5  & 3  &  5  &  2  & 52      & $3.9388 \times 10^{-6}$         & {}                                                     &  {}                                                &  {} \\
 5  & 3 & $\vdots$   & $\vdots$   & $\vdots$             & $\vdots$      &   $\vdots$                            &  {}                                                 &  {} \\
 5 &  4  &  0  &  0  & 0       & 0                                               & 0                                                      &  $7.2300 \times 10^{-5}$                                     & {} \\
 5 &  4  &  1  &  1  & 12       & $3.9739 \times 10^{-5}$        & $3.9739 \times 10^{-5}$                &  {}                                             &  {} \\
 5 &  4  &  2  &  1  & 25       & $9.7145 \times 10^{-6}$        & $2.0233 \times 10^{-5}$                &  {}                                             &  {} \\
 5 &  4  &  2  &  2  & 18       & $1.0518 \times 10^{-5}$        & {}                                                      &  {}                                             &  {} \\
 5 &  4  &  3  &  1  & 38       & $3.5330 \times 10^{-6}$        & $7.9812 \times 10^{-6}$                  &  {}                                             &  {} \\
 5 &  4  &  3  &  2  & 68       & $4.4295 \times 10^{-6}$        & {}                                                      &  {}                                            &  {} \\
 5 &  4  &  3  &  3  & 1         & $1.8712 \times 10^{-8}$        & {}                                                      &  {}                                            &  {} \\
 5 &  4  &  4  &  1  & 68      & $2.0018 \times 10^{-6}$        & $2.8926 \times 10^{-6}$                  &  {}                                            &  {} \\
 5 &  4  &  4  &  2  & 163    & $8.9070 \times 10^{-7}$        & {}                                                      &  {}                                               &  {} \\
 5 &  4  &  4  &  3  & 3        & $8.4996 \times 10^{-11}$      & {}                                                      &  {}                                               &  {} \\
 5 &  4  &  5  &  1  & 112    & $1.2367 \times 10^{-6}$       & $1.4552 \times 10^{-6}$                  &  {}                                                &  {} \\
 5 &  4  &  5  &  2  & 255    & $2.1156 \times 10^{-7}$        & {}                                                     &  {}                                                &  {} \\
 5 &  4  &  5  &  3  & 13      & $6.9973 \times 10^{-9}$        & {}                                                     &  {}                                                &  {} \\
 5 &  4  & $\vdots$   & $\vdots$   & $\vdots$             & $\vdots$      &   $\vdots$                            &  {}                                                 &  {} \\
\br
\end{tabular}
\end{table*}

\begin{table*}
\caption{\label{tab_T0}
The sum rule weights  of the  DSF at $k= k_F$ of the system with $N=L=10$, interaction strength $\gamma=4$ at zero temperature.
The X-sum rule is the total spectral weights of the states under tag X,  and the  $N_\mathrm{p}$-sum rule specifies  the contributions of different pairs of relative p-h excitations.
The tree structure of this sample is shown in Figure~\ref{tree} and the states belonging to the set of tags $(5, 3, 1, 1)$ are demonstrated in Figure~\ref{tag_T0}.
}
\lineup
  \begin{tabular}{c | c | c | c | c c c c c}
  \br
  \multicolumn{4}{c}{\textbf{Tags}} & \textbf{Number of} & \multicolumn{4}{c}{\textbf{Sum Rule}}\\
   \multicolumn{4}{c}{} & \textbf{States}         & \multicolumn{4}{c}{}\\
  \mr
  $P_\mathrm{m}$ & $N_\mathrm{p}$ & $P_\mathrm{l}$ & $N_\mathrm{l}$ &  & $N_\mathrm{l}$-sum rule &  $P_\mathrm{l}$-sum rule & $N_\mathrm{p}$-sum rule & $P_\mathrm{m}$-sum rule \\
  5 &  1  &  0  &  0  & 5       & $0.9759$         & $0.9759$                  & $0.9759$           &  $0.9988$    \\
  5 &  2  &  0  &  0  & 2       & $0.0028$         & $0.0028$                  & $0.0227$           &  {} \\
  5 &  2  &  1  &  1  & 6       & $0.0120$         & $0.0120$                  & {}                       &  {} \\
  5 &  2  &  2  &  1  & 14     & $0.0046$         & $0.0046$                  & {}                       &  {} \\
  5 &  2  &  3  &  1  & 23     & $0.0022$         & $0.0022$                  & {}                       &  {} \\
  5 &  2  &  4  &  1  & 30     & $0.0011$         & $0.0011$                  & {}                       &  {} \\
  5 &  2 & $\vdots$   & $\vdots$   & $\vdots$             & $\vdots$      &   $\vdots$                            &  {}                                               &  {} \\
  5 &  3  &  0  &  0  & 0       & 0                                                            & 0                                                      &  $1.7332 \times 10^{-4}$           & {} \\
  5 &  3  &  1  &  1  & 5       & $7.7379 \times 10^{-5}$                      & $7.7379 \times 10^{-5}$                &  {}                                               &  {} \\
  5 &  3  &  2  &  1  & 16     & $4.7372 \times 10^{-5}$                      & $4.7372 \times 10^{-5}$                &  {}                                               &  {} \\
  5 &  3  &  2  &  2  & 0       & 0                                                            & {}                                                    &  {}                                               &  {} \\
  5 &  3  &  3  &  1  & 42     & $3.0517 \times 10^{-5}$                      & $3.0517 \times 10^{-5}$               &  {}                                               &  {} \\
  5 &  3  &  3  &  2  & 0       & 0                                                            & {}                                                    &  {}                                               &  {} \\
  5 &  3  &  4  &  1  & 79     & $1.8430 \times 10^{-5}$                      & $1.8452 \times 10^{-5}$               &  {}                                               &  {} \\
  5 &  3  &  4  &  2  & 8       & $2.1633 \times 10^{-8}$                      & {}                                                   &  {}                                               &  {} \\
  5 &  3  & $\vdots$   & $\vdots$   & $\vdots$             & $\vdots$      &   $\vdots$                            &  {}                                               &  {} \\
  5 &  $\vdots$  & $\vdots$   & $\vdots$   & $\vdots$             & $\vdots$      &   $\vdots$                            &  {}                                               &  {} \\
\br
\end{tabular}
\end{table*}

In order to illustrate the efficiency of our algorithm, in Figure~\ref{treeT} and Table~\ref{tab} (Figure~\ref{tree} and Table~\ref{tab_T0}) we show a sample of data  for calculating the DSF $S(k=k_\mathrm{F},\omega)$ at the  finite temperature (in ground state).
The temperature and interaction is set by $T=3$ ($T=0$) and $c=4$, while the system size $N=L=10$ for the sake of a clear visibility.
The tree in Figure~\ref{treeT} (Figure~\ref{tree})   obviously shows the data structure of DSF in terms of weights of the sum rule associated with  the quantum numbers  $N_\mathrm{p}$, $P_\mathrm{l}$ and $N_\mathrm{l}$, which  serve as a good criterion for truncation in numerical calculations.
The good saturation of the sum rule confirms the validity of our algorithm for both ground state and finite temperature environments.
We would like to remark that our method  works well  for calculation of other correlation functions, such as one-particle dynamic correlation function at the finite temperatures \cite{Ours_2}.

\section{Results}

\begin{figure}[htbp!]
\centering
\includegraphics[width=1.05\textwidth]{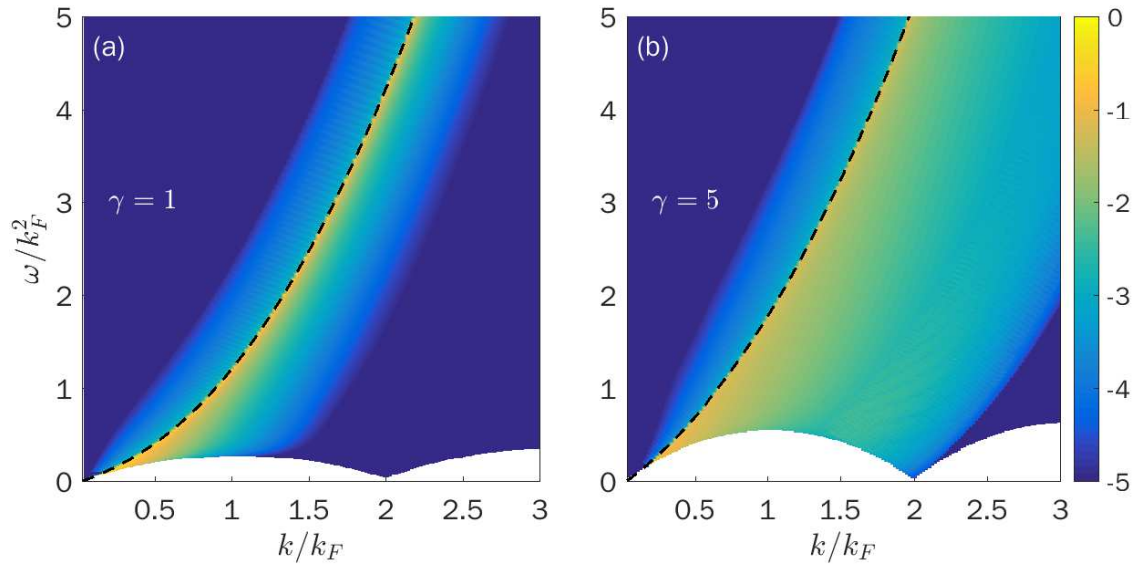}
\caption{(a) and (b) shows  the logarithmic DSF for different interaction strength $\gamma=1$ and $\gamma=5$, respectively.
Without losing generality, we tune the interaction strength parameter $c$ with keeping the particle density intact $n=1$, and here the system size is $N=L=100$.
The momentum and energy are measured in units of Fermi momentum and Fermi energy,  respectively.
The brighter the color, the larger the DSF.
The black dashed curve marks the position of Lieb-I type of dispersion relation, where the DSF diverges in  in thermodynamic limit. }
\label{contour}
\end{figure}

Using the method explained in the above section, we demonstrate the ground state DSFs for the system with different interaction strength $\gamma=1$ and $\gamma=5$ in Figure~\ref{contour}, where the system size is $N=L= 100$.
For the sake of visibility, we plot $\log_{10}(S/L)$ in momentum-energy plane,  where the brighter is the color the larger is the DSF therein.
There are two types of single-particle dispersion relations for this model, i.e. Lieb-I and Lieb-II.
The former (latter) is produced by putting a particle (hole) outside (inside) the Fermi sea which we denote as $\varepsilon_\mathrm{p}$ ($\varepsilon_\mathrm{h}$).
It is blank below the Lieb-II excitation because there is no state bearing energy lower than a single-hole excitation.
The $\varepsilon_\mathrm{p}$- and $\varepsilon_\mathrm{h}$-dispersion relations respectively define the upper and lower thresholds of single-particle spectra.
Above $\varepsilon_\mathrm{p}$, all the signals come from these states generated by multiple pairs of p-h excitations.
It is obvious that the strongest signals of the DSF concentrate along on $\varepsilon_\mathrm{p}(k)$.
In thermodynamic limit, a power-law divergence occurs in the vicinities of the dispersion, which will be studied in more details later.
In comparison of the cases of different interactions, it is apparent that if $\gamma$ vanishes eventually, then only the $\varepsilon_\mathrm{p}$-dispersion survives, and the corresponding DSF finally shrinks into a $\delta$-function; whereas the hole dispersion  $\varepsilon_\mathrm{h}$ disappears as $\gamma$ tends to zero.
In another  word, the interaction changes  the distributions of the spectral weights of  DSF, and the divergence of $\delta$-function is replaced by a power-law  singularity.
To prove the validity of our results given in Figure~\ref{contour}, the values of $f$-sum rule at different momenta $k$ are listed in Table~\ref{sumrule}.

\begin{table}[hbtp]
\caption{$f$-sum rule at momentum $k$ with interaction strength $\gamma=1$ and $\gamma=5$,  respectively. The data come from plot of Figure~\ref{contour}, and the $f$-sum rule shows the precision of our results.}
\label{sumrule}
\centering
\begin{tabular}{c c c c c c c}
\hline
\hline
$k/k_{\rm F}$                & 0.5              & 1.0                 & 1.5                   & 2.0                  & 2.5                & 3.0  \\
\hline
\hline
 $\gamma=1$     & 99.66\%     & 99.22\%        & 98.96\%          & 98.79\%          & 95.30\%       & 95.02\%\\
 \hline
 $\gamma=5$     & 99.74\%     & 99.17\%        & 98.22\%          & 96.98\%          & 95.73\%       & 94.63\%\\
  \hline
\end{tabular}
\end{table}

\begin{figure}[b!]
\centering
\includegraphics[width=0.85\linewidth]{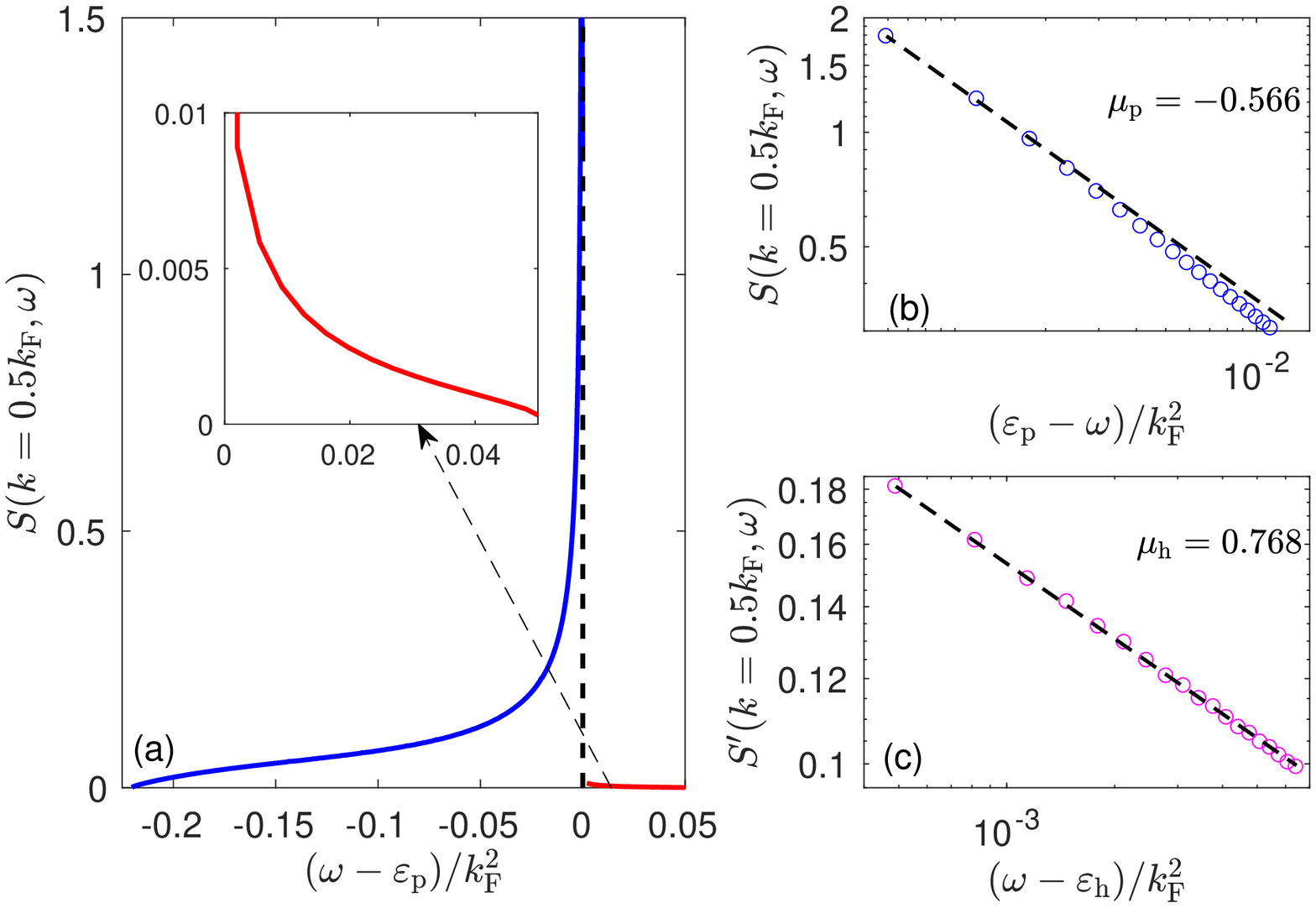}
\caption{(a) shows the lines shape of $S(k,\omega)$ vs $\omega$ with excited momentum $k=0.5 k_F$ and interaction strength $\gamma=1$.
The system size is as large as $N=L=2000$.
The inset shows the rightward of the peak, i.e. the tail of the DSF with large frequency.
(b) and (c) respectively show the power-law behavior of the DSF close to upper and lower spectral thresholds by the log-log coordinate.
The dotted and dashed lines correspond to the original data of calculation  and the extracted exponents, respectively.
By utilizing the nonlinear Tomonaga-Luttinger liquid theory, the exponents on $\varepsilon_\mathrm{p,h}$ are $-0.567$ and $0.810$, agreeing well with our results $-0.566$ and $0.768$, respectively.
}
\label{c1}
\end{figure}

\begin{figure}[hbtp]
\centering
\includegraphics[width=0.85\linewidth]{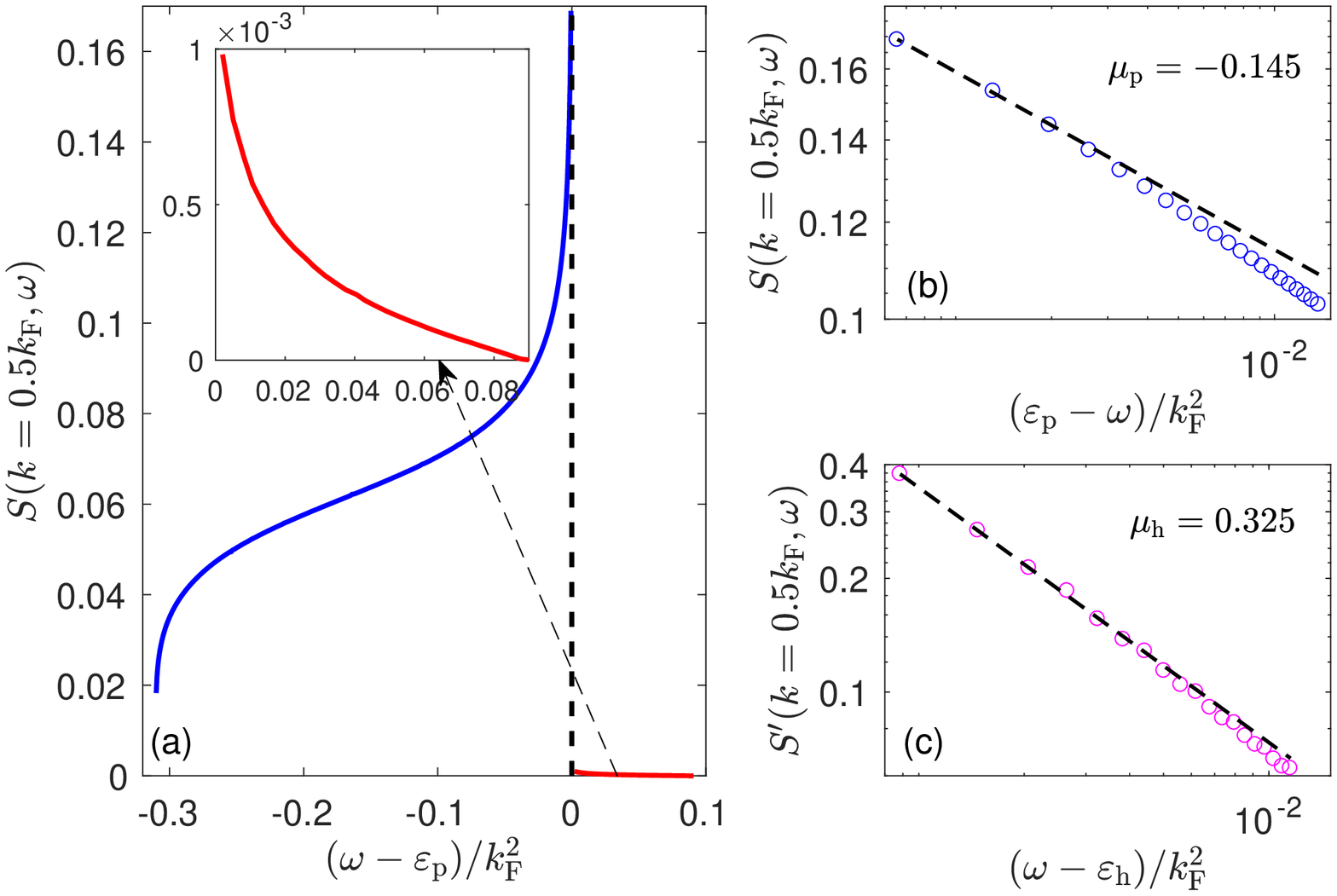}
\caption{(a) shows the lines shape of $S(k,\omega)$ vs $\omega$ with excited momentum $k=0.5 k_F$ and interaction strength $\gamma=1$.
The system size is as large as $N=L=2000$.
The inset shows the rightward of the peak, i.e. the tail of the DSF with large frequency.
(b) and (c) respectively show the power-law behavior of the DSF close to upper and lower spectral thresholds by the log-log coordinate.
The dotted and dashed lines correspond to the original data of calculation  and the extracted exponents, respectively.
By utilizing the nonlinear Tomonaga-Luttinger liquid theory, the exponents on $\varepsilon_\mathrm{p,h}$ are $-0.177$ and $0.202$, agreeing well with our results $-0.145$ and $0.325$, respectively. }
\label{c5}
\end{figure}

The typical line-shapes of DSF for a  given momentum $k=0.5 k_{\rm F}$ of the system at  different interaction strengths $\gamma =1$ and $\gamma =5$ are illustrated  in Figures~\ref{c1} (a) and \ref{c5} (a), where the system size is  taken as $N=L=2000$ in order to capture the threshold singularity.
It clearly shows that for an arbitrary $\gamma$, the signal of the DSF mainly spreads in a certain region  between the particle and hole dispersion lines.
In particular, a sharp  peak in the vicinity of the dispersion  $\varepsilon_\mathrm{p}$ remarkably emerges and it suddenly decreases once the frequency goes over the dispersion line.
In the spectral continuum between $\varepsilon_\mathrm{h}$ and $\varepsilon_\mathrm{p}$, the most weights  of the DSF come from states generated by the $1$-pair of p-h excitation.
In the vicinities  of the two spectral thresholds, those states are in charge of the power-law behavior as well.
As being  mentioned, the DSF at a given momentum is a $\delta$-function in non-interacting limit,  while a flat plateau  shows in TG limit.
A  comparison of Figures~\ref{c1} (a) and \ref{c5} (a) exemplifies the role of interaction in forming the dynamical correlation functions, showing  the evolution of line-shapes between the above two limiting situations.

In Figures~\ref{c1} (b) \& (c) and \ref{c5} (b) \& (c), the behavior of the DSF in the vicinities of spectral thresholds is demonstrated,  showing a power law as following
\begin{equation}\label{FES}
\centering
S(k,\omega) \sim \mathrm{const} + | \omega - \varepsilon_\mathrm{p,h} |^{\mu_\mathrm{p,h}}.
\end{equation}
Here $\mu_\mathrm{p,h}$ stand for the exponents of the DSF  in the vicinities of the Lieb-I and -II dispersion lines, respectively.
The exponent $\mu_\mathrm{p}$ ($\mu_\mathrm{h}$) is negative (or positive), and essentially depends on  $k$ and $\gamma$ \cite{Imambekov2008,Imambekov2009,Imambekov2012,Kitanine2012}.
Such a singularity is caused by the collective behaviour of the interacting particles.
This phenomenon is a reminiscence of Fermi edge singularity (FES) firstly observed in 3D electronic systems \cite{Mahan}.
A simple explanation underlies FES is that an excitation of a deep electron leaves behind a hole in valence band, and this deep hole scatters the electrons in conduction band, resulting in a power-law for spectral function on the threshold \cite{Mahan,Gogolin}.
This puzzle can be figured out by recasting it into a problem of single impurity moving in a cloud of non-interacting electrons.
On account of interaction effect  in 1D, here it evolves into the situation of an impurity in TLL \cite{Gogolin,Balents}.
More specific to the interacting fermions in 1D, one sees  that a moving hole in the Fermi sea scatters the low-lying excited particles.
The physics of this phenomenon is captured  by projecting the system into a three-subbands model, where two subbands are for the low-lying excitations close to both Fermi points and one for the moving hole.
The interaction between the TLL and impurity can be removed by a unitary transformation which is dependent of the phase shifts of excited particles.
Eventually the edge singularity can be extracted by a standard treatment of bosonization \cite{Balents,Gogolin,Imambekov2012}.
The same property for bosonic system and spin chains can be figured out by mapping that system into fermions through the Jordan-Wigner transformation.
In the method of nonlinear TLL, this power-law behavior have been found for both DSF and SF in a variety of 1D quantum systems \cite{Imambekov2008,Imambekov2009,Imambekov2012,Cheianov2008,Pereira2008,Pustilnik2006,Schmidt2010,Kamenev2009}.

For a good  visibility, we use a log-log coordinate in Figures~\ref{c1} (b) \& (c) and \ref{c5} (b) \& (c), and the slops extracted from our data are represented by the black dashed lines.
It should be noted that the treatments of the DSFs for the vicinities of upper and lower thresholds are different.
 In the vicinity of $\varepsilon_\mathrm{p}(k)$, the DSF diverges, hence it is safe to ignore the `const' there.
 For the threshold $\varepsilon_\mathrm{h}(k)$,  the derivative of the DSF with respect to $\omega$, i.e. $S'(k,\omega)$ is employed in order to rule out the influence of the constant background value.
As being  shown in Figures~\ref{c1} (b) \& (c) and \ref{c5} (b) \& (c), the exponents for   $\gamma=1$ ($\gamma=5$) are respectively given by $-0.566$ and $0.768$ ($-0.145$ and $0.325$), in agreement with the nonlinear TLL predictions $-0.567$ and $0.810$ ($-0.177$ and $0.202$) \cite{Imambekov2008,Kitanine2012}.
The advantage of our algorithm is that the quantum numbers  $P_\mathrm{m}$ and $N_\mathrm{p}$, $P_\mathrm{l}$ and $N_\mathrm{l}$ well sort of states in terms of sum rule wights and they serve as a good criterion for making a suitable truncation in the navigation of Hilbert space. The convenient fixing excited momentum $k$ is especially helpful in studying the line-shape of dynamical correlation functions.

\begin{figure}[hbtp]
\centering
\includegraphics[width=0.75\textwidth]{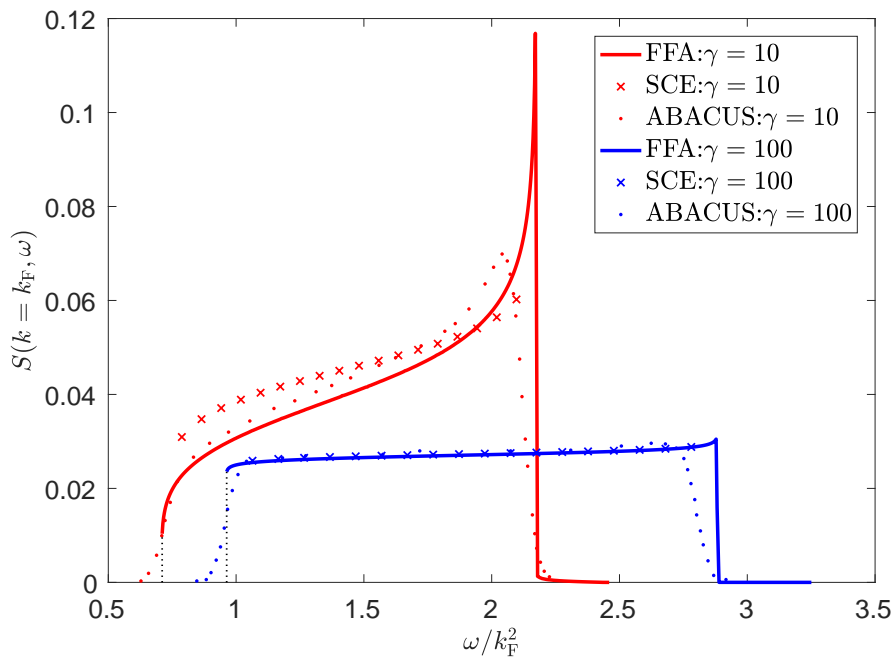}
\caption{The comparison of line-shapes for $S(k_F,\omega)$ obtained from ABACUS, our form factor approach (FFA) and strongly coupling expansion (SCE).
The line, dots and crosses  denote the results of FFA, ABACUS and SCE respectively, while the red and blue indicate the cases of interaction strength $\gamma=10$ and $\gamma=100$ respectively.
The vertical black dashed lines in the left end of DSF specify the Lieb-II excitation energy, below which energy the DSF should vanish.
Note that SCE validates only in the continuum between $\varepsilon_\mathrm{h}$ and $\varepsilon_\mathrm{p}$.
The system size to implement FFA is set as $N=L=1000$, and results of ABACUS come from \cite{Caux2006} with system size $N=L=100$.}
\label{comparsion}
\end{figure}

Our algorithm is quite different from the one developed in  ABACUS  \cite{Caux2009}.
In Figure~\ref{comparsion} we compare results from our form factor approach (FFA)  with the that obtained from the strongly coupling expansion (SCE) method  \cite{Brand2005} and the  ABACUS \cite{Caux2006}.
The SCE is only valid in the continuum between $\varepsilon_\mathrm{h}$ and $\varepsilon_\mathrm{p}$, and expressed as following
\begin{equation}\label{SCE}
\frac{S(k,\omega)}{L}= \frac{1}{4\pi k} \left( 1+\frac{8}{\gamma} \right) + \frac{1}{2\pi \gamma k_{\rm F}}
\ln \frac{\omega-\varepsilon_\mathrm{h}^2}{\varepsilon_\mathrm{p}^2 - \omega^2} + o \left( \frac{1}{\gamma^2}\right).
\end{equation}
In the case of $\gamma=100$, we observe that  the  three results in the continuum agree with each other, however, the agreements  break down in the neighborhood of thresholds.
This is partly because the limitation of ABACUS algorithm and thus a small system size here $N = L = 100$, and partly because a Gaussian is used to replace the Dirac-delta
function in Equation~\ref{spectral_form}.
The latter is to smoothen the line-shape of dynamical correlation functions in the continuum when the system size is not large enough, nevertheless at
the cost of blurring their threshold behavior \cite{note2}.
For the case of $\gamma=10$, as being expected, the performance of SCE is not good, whereas  the results obtained from the ABACUS and  that FFA basically agree with each other  except  in the vicinities of spectral thresholds.

\section{Conclusion}

We have introduced  `relative excitations' characterized by the quantum numbers $N_\mathrm{p}$, $P_\mathrm{l}$ and $N_\mathrm{l}$ for  our  new  algorithm to precisely calculate the dynamical correlation functions over an arbitrary state, i.e. either the ground state or the equilibrium states at finite temperatures.
In our algorithm, it is easy to fix the excited momentum by the tag of quantum number $P_\mathrm{m}$, so is the truncation of the evaluation with the help of quantum numbers $N_\mathrm{p}$, $P_\mathrm{l}$ and $N_\mathrm{l}$.
Such a set of tags is very convenient to  implement the calculation of dynamical correlation functions at a very large scale.
Using this newly developed algorithm, we have presented the exact DSF of Lieb-Liniger model by means of form factor approach.
We have explicitly obtained  the power-law behavior of the DSF in the vicinities of single-particle spectral thresholds at a many-body level  $N=2000$.
While  the typical line-shape of $S(k,\omega)$ vs $\omega$ with given momentum has been demonstrated as well.
On top of the full picture of line-shape, the corresponding exponents of edge singularity are extracted and further confirm  the validity of  nonlinear TLL theory besides \cite{Kitanine2012}.
Furthermore, we have  compared our results with that obtained from  the ABACUS and strongly coupling expansion SCE, and observed their  discrepancies from our calculations and the nonlinear TLL theory, indicating that the Gaussian used in the ABACUS  is unnecessary at a large system size.
Our work offers a reliable approach  to  the correlation properties merely emergent in thermodynamic limit, and manifests the form factor approach itself as a benchmark of the state-of-art  ultra-cold atom experiments in near future \cite{Exp,Guan2022,Guan2013}.

\ack
This work is supported by National Natural Science Foundation of China Grants Nos. 12104372, 12047511, 12134015, 11874393 and 12121004.
SC acknowledges the computational resources from the Beijing Computational Science Research Center.

\section*{Appendix}
\appendix
\setcounter{section}{1}

\subsection*{Form factor}
Here we present  the necessary formulas for our numeric evaluation.
The theoretical derivation of the form factor was given in  \cite{Slavnov}.
By using form factor, the DSF is expressed in terms of rapidities as following
\begin{equation}\label{dsf}
\fl S(k,\omega) = L^2 \sum_{ \{ \mu\} } k^2 \prod_{j,k}^{N} \frac{\lambda_{jk}^2 + c^2}{\left( \mu_j - \lambda_k \right)^2} \frac{\left\| \det_N U(\{\mu\},\{\lambda\}) \right\|^2}{\left\| V_p^+ - V_p^- \right\|^2  \left\| \{\mu \} \right\|^2 \left\| \{\lambda \} \right\|^2} \delta_{k,P_{\mu,\lambda,}} \delta \left( \omega - E_{\mu,\lambda} \right)
\end{equation}
where for simplicity we denote $\nu_{jk} \equiv \nu_j - \nu_k$, and $O_{\mu,\lambda} \equiv O_\mu - O_\lambda$ with $O = E$ or $P$. In above equation, $p$ can be any integer $p\in [1,N]$, and we usually set it as $N$. The norm square of eigenstate $ \| \{ \nu \} \|^2$ reads
\begin{equation}
 \| \{ \nu \} \|^2 =  c^N \prod_{j>k}^{N} \frac{\nu_{jk}^2 + c^2}{\nu_{jk}^2} \det_N \mathcal{G}(\{ \nu\})
\end{equation}
where the entry of matrix $\mathcal{G}(\{\nu\}) $ is expressed by
\begin{equation}
\mathcal{G}_{jk}(\{\nu\}) = \delta_{jk} \left[ L + \sum_{a=1}^{N} K (\lambda_j,\lambda_a)  \right] - K (\lambda_j,\lambda_k)
\end{equation}
with kernel function
\begin{equation}\label{kernel}
K (x,y) = \frac{2c}{(x-y)^2 + c^2}.
\end{equation}
The matrix $V^\pm$ and $U\left(\{\mu\},\{\lambda\}\right)$ are respectively defined by following entries
\begin{equation}
  V_j^\pm  = \frac{\prod_{a=1}^{N-1} \left( \mu_a - \lambda_j \pm \rmi c \right)}{\prod_{b=1}^{N} \left( \lambda_b - \lambda_j \pm \rmi c \right)},
\end{equation}
and
\begin{equation}
\fl U_{jk}\left(\{\mu\},\{\lambda\}\right)  = \delta_{jk} \frac{\left( V_j^+ - V_j^- \right)}{\rmi }
  + \frac{\prod_{a}^{N-1} \left( \mu_a - \lambda_j \right)}{\prod_{b \neq j}^N \left( \lambda_b -\lambda_j \right)} \left[ K(\lambda_j,\lambda_k) - K(\lambda_N,\lambda_k) \right].
\end{equation}

\subsection*{Exponents determined by the  nonlinear Tomonaga-Luttinger liquid theory }
We explain  how to carry out the calculation of exponents by using the nonlinear Tomonaga-Luttinger liquid theory.
The interested readers may refer to \cite{Imambekov2008} for more details.
\begin{equation}
S(k,\omega) \sim \mathrm{const} + | \omega - \varepsilon_\mathrm{p,h} |^{\mu_\mathrm{p,h}}.
\end{equation}
In order to obtain above exponents, we need to solve two integral equations. One is the shift function \cite{QISM}
\begin{equation}\label{shift}
F_B (\nu|\lambda) = \frac{\pi+\theta(\nu-\lambda)}{2\pi} +\frac{1}{2 \pi} \int_{-q}^{q} \rmd \mu\,  K(\nu,\mu) F_B(\mu|\lambda)
\end{equation}
where $\theta(x)=2\arctan(x/c)$ and $K(x,y)$ is defined by equation~(\ref{kernel}).
Note that $q>0$ is the cut-off of rapidity for ground state in thermodynamic limit.
The other one is for the dressed momentum, i.e. the change of total momentum when adding a particle (hole) with rapidity $\lambda >q$ ($|\lambda|<q$) to the ground state.
This change $k(\lambda)$ is expressed by
\begin{equation}\label{momentum_shift}
k(\lambda) = \pm  \left(\lambda - \pi n + \int_{-q}^{q} \rmd \nu\, \theta(\lambda-\nu) \rho(\nu)  \right)
\end{equation}
where $\pm$ is to specify adding a particle or hole and $\rho(x)$ is the distribution of rapidity in thermodynamic limit, governed by following integral equation
\begin{equation}\label{rho}
\rho(\lambda)= \frac{1}{2\pi} + \frac{1}{2\pi} \int_{-q}^{q} \rmd \mu K(\lambda,\mu) \rho(\mu).
\end{equation}
The exponent reads
\begin{equation}\label{exponent}
\mu =  \frac{1}{2} \left( \frac{1}{\sqrt{K}} + \frac{\delta_+ - \delta_-}{2\pi} \right)^2 + \frac{1}{2}  \left( \frac{\delta_+ + \delta_-}{2\pi} \right)^2 - 1
\end{equation}\label{exponents}
where $\delta_\pm = 2\pi F_B(\pm q, \lambda)$ and $K$ is the Luttinger parameter.
Accordingly, one may numerically solve above integral equations to obtain $\mu_{\rm p}(k)$ ($\mu_{\rm h}(k)$).
This exponent describes the power-law behavior of the DSF on the upper (lower) threshold, which dispersion is generated by adding 1-particle  (1-hole) to the ground state,
 and thus the sign in Equation~(\ref{momentum_shift}) is positive (negative).
At first we solve the corresponding $\lambda_{\rm p}$ ($\lambda_{\rm h}$) by using Equation~(\ref{momentum_shift}), and then substitute this $\lambda_{\rm p}$ ($\lambda_{\rm h}$) into the shift function $F_B(\mu|\lambda_{\rm p})$ ($F_B(\mu|\lambda_{\rm h})$).
 With the help of Equation~(\ref{shift}), $\delta_\pm(k)$ is obtained, so is the exponent $\mu_{\rm p}(k)$ ($\mu_{\rm h}(k)$).

\end{document}